\documentclass[apj]{emulateapj}

\shorttitle{3D GRB simulations}
 
\shortauthors{Lopez-Camara et al.}

\begin{document}

\title{Three-dimensional AMR simulations of long-duration Gamma-Ray
  Burst jets inside massive progenitor stars}

\author{D. L\'opez-C\'amara\altaffilmark{1, *}, Brian J.
  Morsony\altaffilmark{2}, Mitchell C. Begelman\altaffilmark{3, 4},
  Davide Lazzati\altaffilmark{1}} 
  
\altaffiltext{1}{Department of Physics, NC State University, 2401
  Stinson Drive, Raleigh, NC 27695-8202, USA}

\altaffiltext{2}{Department of Astronomy, University of
  Wisconsin-Madison, 2535 Sterling Hall, 475 N. Charter Street,
  Madison WI 53706-1582, USA}

\altaffiltext{3}{JILA, University of Colorado, 440 UCB, Boulder, CO
  80309-0440, USA}

\altaffiltext{4}{University of Colorado, Department of Astrophysical
  and Planetary Sciences, 389 UCB, Boulder, CO 80309-0389, USA}

\altaffiltext{*}{dlopezc@ncsu.edu}
  
\begin{abstract}

  We present the results of special relativistic, adaptive mesh
  refinement, 3D simulations of gamma-ray burst jets expanding inside
  a realistic stellar progenitor. Our simulations confirm that
  relativistic jets can propagate and break out of the progenitor star
  while remaining relativistic. This result is independent of the
  resolution, even though the amount of turbulence and variability
  observed in the simulations is greater at higher resolutions.  We
  find that the propagation of the jet head inside the progenitor star
  is slightly faster in 3D simulations compared to 2D ones at the same
  resolution. This behavior seems to be due to the fact that the jet
  head in 3D simulations can wobble around the jet axis, finding the
  spot of least resistance to proceed. Most of the average jet
  properties, such as density, pressure, and Lorentz factor, are only
  marginally affected by the dimensionality of the simulations and
  therefore results from 2D simulations can be considered reliable.
 
\end{abstract}

\keywords{gamma-ray bursts: general --- hydrodynamics --- supernovae: general}

\section{Introduction}\label{sec:intro}

Long-duration gamma-ray bursts (GRBs) are produced by collimated
relativistic outflows \citep{sari99} ejected in the core of massive
stars at the end of their evolution \citep{w93, hjorth03, s03, wb06}.
Since their relativistic outflows have to propagate through their
progenitor star material and exit the star before producing the
gamma-ray photons, an outstanding issue with this scenario is to
understand the mechanisms that prevent the entrainment of baryons in
the light, hot jet \citep{mw99, aloy00}. 

On the other hand, even if the jet-star interaction cannot slow down
the jet, it has a strong impact on its dynamics \citep{mor07} and can
supply enough energy to explode the star as a supernova
\citep{khokhlov99, mwh01, wheeler02, maeda03, laz12}.  In most cases,
the study of the jet-star interaction has been performed numerically,
with analytic models used only for guidance \citep{aloy02, gomez04,
  mor07, matzner03, brom11}.  Even so, studying the propagation of a
relativistic outflow that is continuously shocked by a much denser
environment is not trivial since the length-scale of features in the
relativistic material is typically $\sim R/\Gamma$ and therefore a
large dynamical range is involved. When possible, adaptive mesh
refinement (AMR) codes have been adopted \citep{mor07, mor10, laz09,
  laz10, laz11b, nag11}, and the simulations have been limited to two
dimensions \citep{mw99, aloy00, mwh01, zwm03, miz06, mor07, mor10,
  laz09, laz10, laz11b, miz09, nag11}. These studies have shown that
even though the jet material is relativistic, the jet-head propagates
sub-relativistically inside the star, thereby allowing causal contact
between the bow shock at the head of the jet and the star. The shocked
star material therefore drains at the sides of the jet producing a hot
cocoon \citep{rr02, laz05} instead of being entrained in the jet.

Two dimensional (2D) simulations can provide important answers to the
outstanding questions listed above. However, they are plagued by
artifacts due to the presence of a symmetry axis in the center of the
jet. First, a plug of dense material accumulates in front of the jet
head, slowing down its propagation and creating plumes of hot plasma
at wide angles (see Figure~1 in \citet{laz10} for an example).
Second, recollimation shocks coming from the sides of the jet bounce
strongly off the jet axis in 2D simulations, while they could
dissipate more efficiently in a simulation at the natural
dimensionality.  Finally, the role of turbulence and instabilities
cannot be properly explored in 2D simulations. \citet{wang08} found
that in some cases a three dimensional (3D) relativistic jet would
break apart and not be able produce a successful GRB (while in 2D it
would produce a successful GRB).

While 3D simulations of GRB jets have been attempted in the past
\citep{zwh04}, they were performed with a fixed grid code, casting
doubt on their capability to resolve the required small scales.  A 3D
test-case with AMR was presented by \citet{wang08}, but since the
jet-progenitor evolution varied drastically as a function of the
numerical resolution (unlike our study), not much could be inferred
from their study.  Thus, in this paper we present, for the first time,
3D adaptive mesh refinement (AMR) simulations of GRB jets crossing a
pre-SN progenitor and then flowing through the interstellar medium.

This paper is organized as follows. We first describe the physics,
initial setup, and the numerical simulations in
Section~\ref{sec:input}, followed by our results and discussion in
Section~\ref{sec:results}. Conclusions are given in
Section~\ref{sec:conc}.

\section{Physics, initial setup and simulations}\label{sec:input}
\subsection{Physics and initial setup}\label{sec:phys&initsetup}

As what now seems to be the generic model used for long GRBs
\citep[for example]{mor07, mor10, laz09, laz11a, laz11b, laz12, lc09,
  lc10, lind10, lind12, nag11}, we consider the one-dimensional (1D)
pre-supernova 16TI model from \citet{wh06} as our initial stellar
configuration.  Initially (in the zero-age main sequence) model 16TI
is a 16$\,M_\odot$ Wolf-Rayet star with 0.01$\,Z_\odot$ metallicity,
and $3.3 \times 10^{52}$~erg~s equatorial angular momentum. The final
outcome of such model is a pre-SN progenitor with 13.95$\,M_\odot$ and
nearly half the size of the sun ($R_0 = 4.1 \times
10^{10}$~cm). Assuming spherical symmetry, the 1D density and pressure
profiles were mapped onto a 3D configuration that we assumed to be
initially without rotation. The internal energy and the temperature
were calculated assuming a relativistic polytropic equation of state
($\gamma$=4/3).  The pre-SN progenitor was immersed in an interstellar
medium (ISM) with constant density
($\rho_{\rm{{ism}}}=10^{-10}$~g~cm$^{-3}$). Even though a wind
environment would probably be more appropriate, we note that within
the size of our simulation the dynamical role of the ambient medium is
negligible and the results are therefore insensitive to the chosen
ambient medium profile.

A relativistic jet commencing its flow at the center of the pre-SN
progenitor was imposed at all times as a boundary inflow condition. The jet
was launched at the center of the star (in fact slightly above it),
flowing upwards in the polar direction (x=z=0,
y=R$_{\rm{{i}}}=$10$^9$cm). The imposed jet had a half-opening angle
of $\theta_0$=10$^{{o}}$, a constant luminosity of L$_0=5.33
\times$10$^{50}$~erg~s$^{-1}$, an initial Lorentz Factor of
$\Gamma_{\rm{{0}}}$=5, and a ratio of internal over rest-mass energy
equal to $\eta_0$=80 \citep{mor07, mor10, laz09}. In order to break
the 2D axis symmetry, the jet was slightly asymmetric. For the latter, we set
the jet with a 1\% density and pressure asymmetry on either side of a
line in the XZ plane 40 degrees from the X axis. Differently from
\citet{wang08} (3D numerical study in which a two dimensional
symmetrical initial setup was assumed) our initial setup resembles
that from model 3A in \citet{zwh04} enhanced with a small perturbation
in the jet.

\subsection{Numerical simulations}\label{sec:sims}

In order to follow the temporal evolution of our initial setup, we
solved the 3D gas-dynamic equations using the FLASH code (version 2.5) in cartesian coordinates
\citep{fryx00}. The simulation domain covered the top half of the
pre-SN progenitor star as well as the ISM it is immersed in (see for
example panel $a$ from Figure~\ref{fig1}). The boundaries were set at
y$_{\rm{min}}$=10$^9$cm, y$_{\rm{ {max}}}$=2.4$\times 10^{11}$~cm,
x$_{\rm{ {max}}}$=-x$_{\rm{ {min}}}$=6$\times 10^{10}$~cm, and
z$_{\rm{ {max}}}$=-z$_{\rm{ {min}}}$=6$\times 10^{10}$~cm. Only the
equatorial plane (y=y$_{\rm{ {min}}}$) was set with a reflective
boundary condition, all the other boundaries were set with
transmission conditions.  
We used a 10-level binary adaptive grid with square-shaped pixels ($\Delta{x}=\Delta{y}=\Delta{z} \equiv \Delta$). The highest refinement level (also referred to as the finest
resolution level) were accessible only at the core of the pre-SN star
were the jet is injected and initially propagates. Moving away from
the stellar core, the maximum level of refinement was progressively
decreased. In practice, the base of the jet had the finest resolution
at all times and the three next finest levels followed the jet (and
the polar part of the cocoon) as it drilled through the progenitor.

Two set of simulations with a different value of $\Delta$ were
performed. We will refer to the the hight resolution model as ``HR'',
and the low resolution as ``LR''. The HR model had the finest
resolution (covering the core of the star at all times) equal to
$\Delta=3.125 \times 10^{7}$~cm, the jet for this case was followed
with a resolution of at least $\Delta=1.25 \times 10^{8}$~cm. The LR
model had the same setup but the value of the the finest resolution
was set equal to $\Delta=6.25 \times 10^{7}$~cm, and the jet was
followed with a resolution of at least $\Delta=2.5 \times
10^{8}$~cm. The resolution with which we follow the jet is comparable
to that from the 3D collapsar study of \citet{zwh04} (where the
maximum resolution was $\Delta \sim$10$^{8}$~cm), and to the most
recent 3D GRB jet study from \citet{wang08} (where $\Delta=7 \times
10^{7}$~cm). The resolution with which we resolve the core of the star
is comparable to that from previous 2D GRB jet numerical studies
\citep{zwm03, miz06, mor07, nag11}.  In order to understand the
three-dimensional effects properly, we also ran an extra two
dimensional model. 

Differently from the 3D simulation, the 2D run was performed in
cylindrical coordinates, the polar axis being coincident with the jet
axis. The 2D model had an initial configuration akin to the XY and
the ZY planes of the 3D model. The 2D model had the same input physics
and resolution as that of the 3D HR model. A summary with the
differences between the numerical models is shown in Table 1.

\begin{table}[htdp]
\tabletypesize{\footnotesize}
\caption{Model characteristics}
\begin{center}
\begin{tabular}{ccc}
\hline
Model & $\Delta$ in core & $\Delta$ in jet \\
           & ($\times 10^{7}$~cm) & ($\times 10^{8}$~cm) \\
\hline
3D LR & 6.250 & 2.50 \\
3D HR & 3.125 & 1.25 \\
2D HR & 3.125 & 1.25 \\
\hline
\end{tabular}
\end{center}
\label{default}
\end{table}

\section{Results and Discussion}\label{sec:results}
\subsection{Global morphology. }\label{sec:morph}

\begin{figure*}[ht!]
\begin{center} 
\includegraphics*[width=0.7\textwidth]{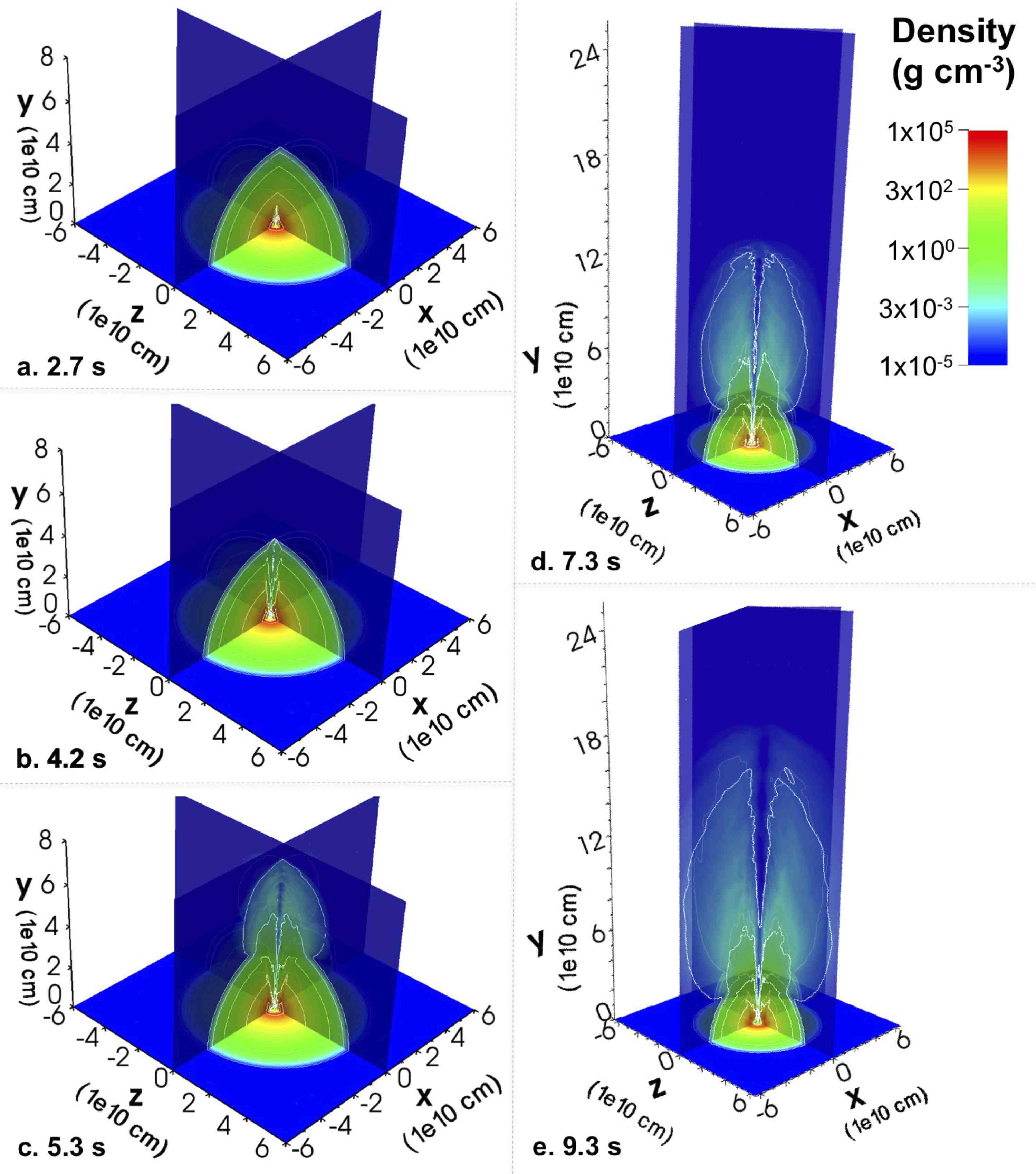}
\caption{Density stratification maps (g~cm$^{-3}$) for different
  timeframes (a. 2.7~s; b. 4.2~s; c. 5.3~s; d. 7.3~s; e. 9.3~s) for
  model 3D LR.  The isocontour levels correspond to:
  10$^{4}$~g~cm$^{-3}$; 10$^{2}$~g~cm$^{-3}$; 1~g~cm$^{-3}$;
  10$^{-2}$~g~cm$^{-3}$; 10$^{-4}$~g~cm$^{-3}$. In order to better
  visualize the internal structure in the pre-SN, the minimum value in
  all the density stratification plots was set to
  10$^{-5}$g~cm$^{-3}$. A movie of this figure is available in http://www4.ncsu.edu/$\sim$dlopez/Simulations$\underline{\hspace{0.2cm}}$(published).html}
    \label{fig1}
\end{center} 
\end{figure*}

\begin{figure*}[ht!]
\begin{center} 
\includegraphics*[width=0.7\textwidth]{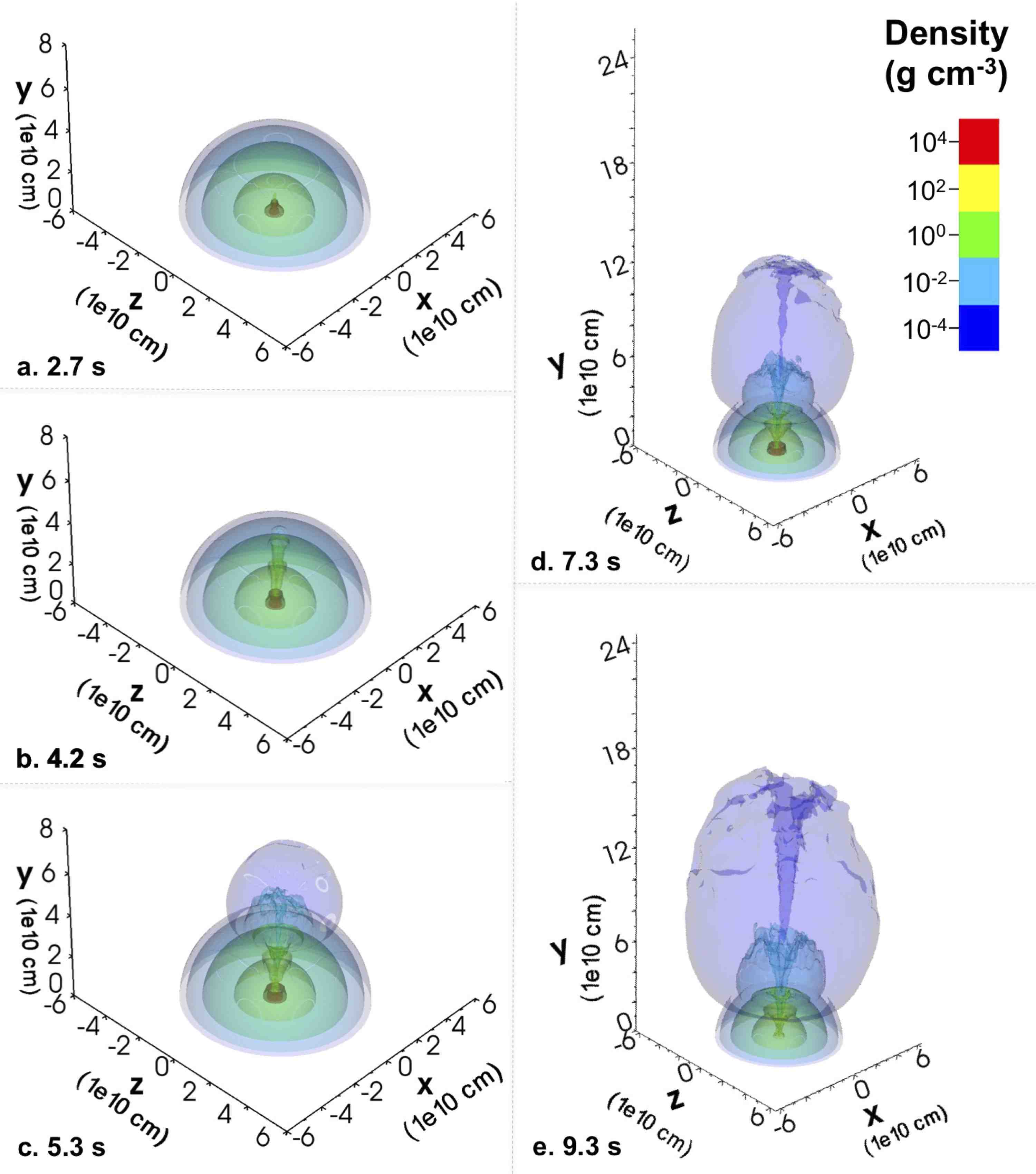}
\caption{Density stratification map for different isocontour levels
  (red=$10^{4}$~g~cm$^{-3}$; yellow=$10^{2}$~g~cm$^{-3}$;
  green=1~g~cm$^{-3}$; cyan=$10^{-2}$~g~cm$^{-3}$;
  blue=$10^{-4}$~g~cm$^{-3}$) for model 3D LR. The timeframes are the
  same as those indicated in Figure~\ref{fig1}. A movie of this figure is available in http://www4.ncsu.edu/$\sim$dlopez/Simulations$\underline{\hspace{0.2cm}}$(published).html}
    \label{fig2}
\end{center} 
\end{figure*}

In Figure~\ref{fig1} we show the density stratification maps for the
XZ, XY, and ZY planes for the 3D LR model. Each panel shows a
different timeframe: a. t$_a = 2.7$~s; b. t$_b = 4.2$~s; c. t$_c =
5.3$~s; d. t$_d = 7.3$~s; and e. t$_e = 9.3$~s. These panels are
arranged to illustrate the jet-progenitor-ISM temporal evolution
(animations of the density stratification map, Lorentz factor, radial
density, radial Lorentz factor, and Schlieren map in the XY plane, are
linked to the online version of this manuscript). The morphology of
our system is divided into two main phases: when the jet moving inside
the progenitor and when the jet has broken out of the star and
interacts with the interstellar medium. Such temporal evolution is
consistent with what has already been seen in previous numerical
studies \citep{zwm03, mor07}. Superimposed on the density
stratification in Figure~\ref{fig1}, we show the isocontour levels
corresponding to 10$^{-4}$, 10$^{-2}$, 1, 10$^{2}$, and 10$^{4}$ (all
in g~cm$^{-3}$), these isocontour levels are shown in
Figure~\ref{fig2}.

The t$_{\rm{ {bo}}}$=4.2~s breakout time is similar to (but somewhat
shorter than) that already seen in previous collapsar studies
\citep{zwm03, zwh04, mor07, mor10}. Depending on the progenitor that
one chooses, and the particular characteristics of the jet, it takes 5
to 10~s to cross the stellar envelope. Compared to power-law stellar
models, the models from \citet{wh06} are more compact and dense and it
takes less time for the jet to cross the realistic progenitors
\citep{miz06}. Our t$_{\rm{ {bo}}}$ is very similar to the breakout
time computed with the analytical model from \citet{brom11}. Still, it
must be stated that since the jet in our numerical simulations is
launched at an inner radius which is at least 10$^4$ times the
gravitational radius (R$_{\rm{{i}}} \sim$ $10^{4}$R$_g$ for a
1.4M$_\odot$ black hole), the jet from the simulations is somewhat
wider than that from the analytical model and thus it propagates
slower.  The t$_{\rm{ {bo}}}$ for our study implies that the average
propagation velocity of the jet inside the star is $\sim0.32c$. The
jet, composed of low density material, has its initial opening angle
reduced by relativistic hydrodynamic collimation effects.

Once the jet crosses the stellar envelope and breaks out of the
surface, the cocoon (which surrounds the jet and is present since its formation) expands through the ISM \citep{rr02, laz05}, differently from when the
jet is drilling through the progenitor when the cocoon is bound inside
the star and close to the jet. When the jet breaks out of progenitor
it becomes uncollimated and the cocoon moves out in the polar
direction (moving parallel to the jet), also expanding sideways on top
of the stellar surface. Such spreading (see panel c, d, and e from
Figures~\ref{fig1}-\ref{fig2}) was predicted by the analytic solution
from \citet{brom11}.  By this time not only does the cocoon present
zones where variability is clearly present, but also the jet presents
turbulent-like structures. The variability in the cocoon is due to the
fact that the jet-cocoon system is at least five orders of magnitude
denser than the surrounding ISM. Hence, any instability that forms on
the cocoon's boundary or that travels upwind from the jet into the
cocoon is not dissipated. Due to the location of the outer boundaries,
we are not able to follow the jet-cocoon-ISM system entirely after
approximately 10~s. By this time the cocoon has crossed the outer
boundaries (the jet crosses the top boundary at approximately 13~s).
Also, it must be noted that as time passes the inner isocontour
($\rho$=10$^{4}$~cm~g$^{-3}$) disappears. This is due to the reverse
shock, which is expanding and pushing the dense material
outwards. Such behavior has already been seen in the study of
\citet{laz10}.

\subsection{Symmetry loss}\label{sec:simloss}

\begin{figure}[ht!]
\begin{center} 
\includegraphics*[width=0.45\textwidth]{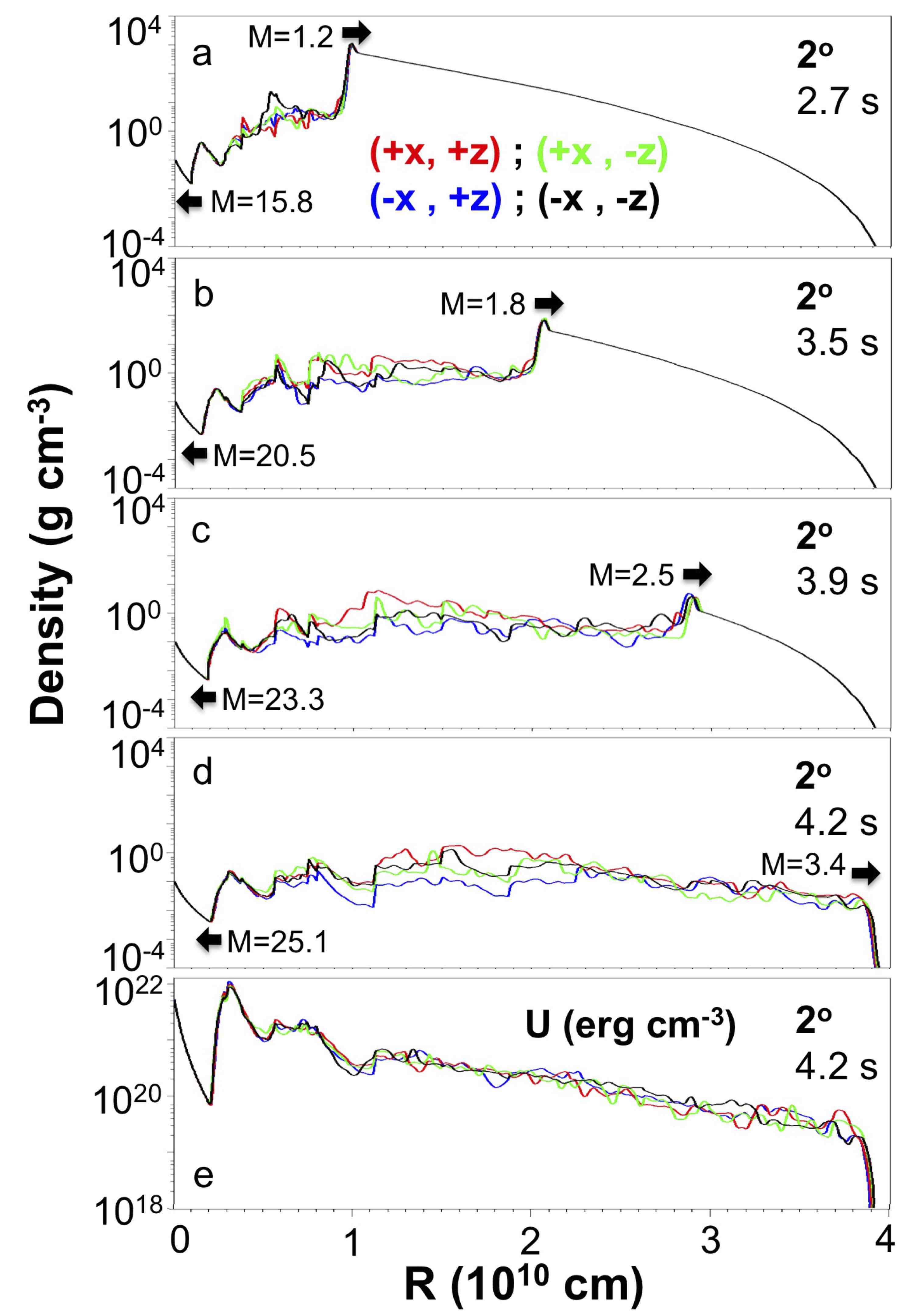}
\caption{Radial profiles for different 2$^{ {o}}$ paths and
  times for model 3D LR. Panel a through d show different radial density profiles
  (g~cm$^{-3}$), panel e shows different energy density profiles
  (erg~cm$^{-3}$). Each of the radial paths commences in the origin
  and runs through different quadrants: (+X,+Z) quadrant (red line);
  (+X,-Z) quadrant (green line); (-X,+Z) quadrant (blue line); (-X,-Z)
  quadrant (black line). Each panel corresponds to a different
  timeframe: a. 2.7~s, b. 3.5~s, c. 3.9~s, d. and e. 4.2~s. The
  forward and reverse shock Mach number for each timeframe are also
  indicated. A movie of this figure is available in http://www4.ncsu.edu/$\sim$dlopez/Simulations$\underline{\hspace{0.2cm}}$(published).html}
    \label{fig3}
    \end{center} 
\end{figure}

To understand when the cylindrical symmetry is broken, in
Figure~\ref{fig3} we plot the radial density distribution as well as
the energy density (U, in erg~cm$^{-3}$) for four different paths
which move along a cone of 2$^{ {o}}$ half-opening angle (with its
origin set at x=y=z=0) for model 3D LR. One of these paths moves
radially (R) along the ``(+X,+Z)'' quadrant; another moves
in the ``(+X,-Z)'' quadrant; another in the ``(-X,+Z)'' quadrant, and
finally a path which moves along the ``(-X,-Z)'' quadrant.

Consistently with previous 2D and 3D collapsar simulations
\citep{zwm03, zwh04, miz06,nag11}, we see that as the jet drills
through the stellar envelope a complex shock system forms,
characterized by a forward and a reverse shock at the head of the jet
and by and a series of conical recollimation shocks. The first
recollimation shock, visible almost at the base of the jet, seems to
be static (at R$\sim$2$\times$10$^9$~cm), but this is due to the fact
that it moves at relativistic speed in the rest-frame. Since the study of the shock structure is not
our goal, we do not focus on the nature of these shocks, nor do we
need to know where the contact discontinuity is set. For the sake of
our study all we need to be able to discern is the stellar and jet
material that has and has not been shocked.

\begin{figure}[ht!]
\begin{center} 
\includegraphics*[width=0.45\textwidth]{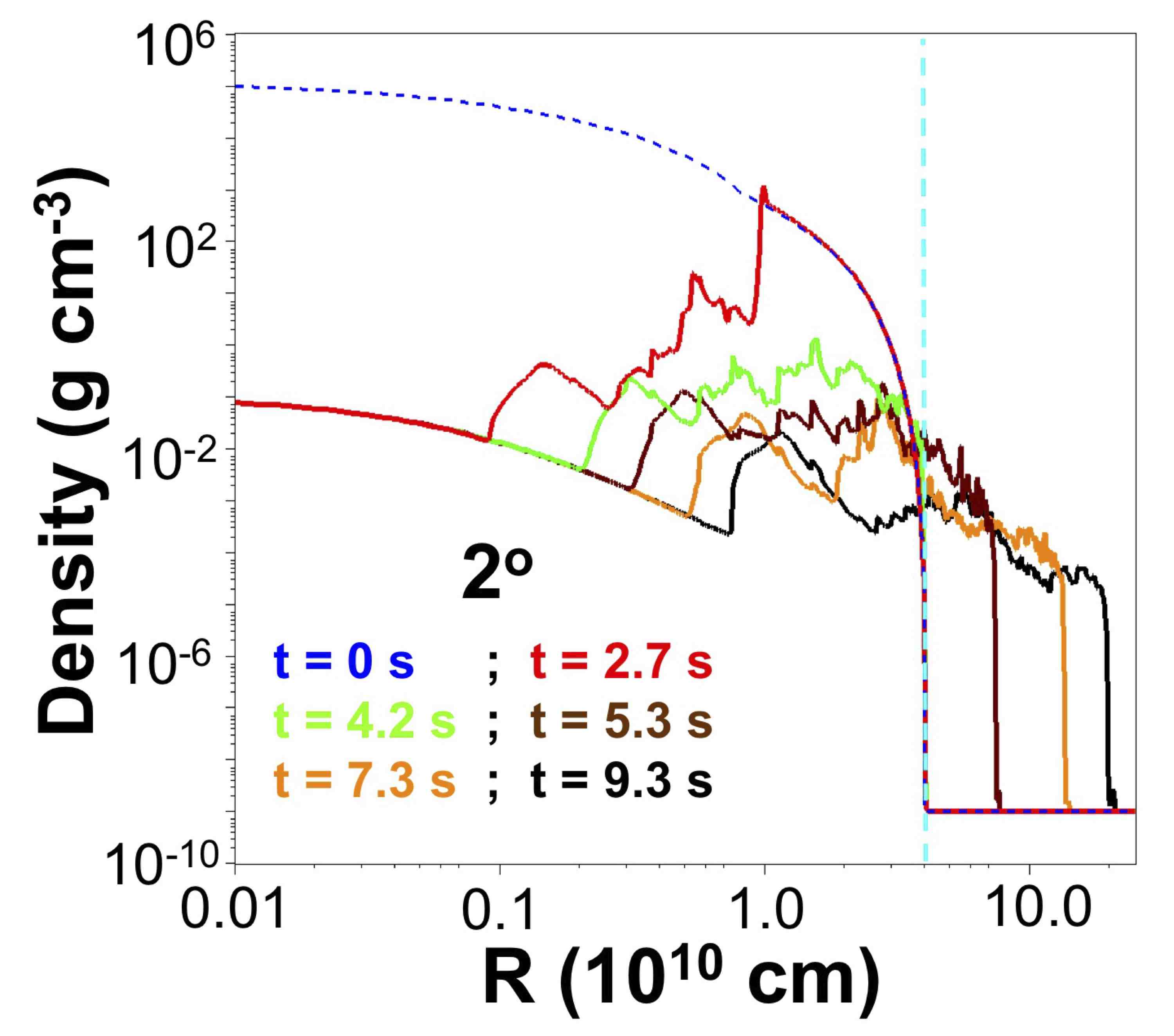}
\caption{Time evolution (0~s; 5.3~s; 7.3~s; 9.3~s) for the 2$^{ {o}}$
  radial density profile (g~cm$^{-3}$) from the (+X,+Z) quadrant
  (black line in Figure~\ref{fig3}). The stellar surface is indicated
  by the cyan dashed line. A movie of this figure is available in http://www4.ncsu.edu/$\sim$dlopez/Simulations$\underline{\hspace{0.2cm}}$(published).html}
    \label{fig4}
    \end{center} 
\end{figure}

Specifically, the regions which we will be addressing to in the rest
of the discussion will be the shocked (SJ) and unshocked (UJ) parts of
the jet. The UJ material maintains its initial density profile, while
the SJ material breaks the symmetry in the 3D numerical
simulations. The density profile can vary up to two orders of
magnitude for different locations at the same distance from the
progenitor center; on the other hand, the UJ varies less than an order
of magnitude.
As the jet crosses through the progenitor star its density decreases
as a function of time (see Figure~\ref{fig4}). Before the jet breaks
out from the stellar surface, the density profile inside the
progenitor follows a quasi-constant profile which for t$_{\rm{ {bo}}}$
is $\sim$10$^{-1}$~g~cm$^{-3}$.  Then, when the jet breaks out of the
stellar surface, it recovers a decaying radial density profile that
for 9.3~s reaches density values as low as 10$^{-5}$~g~cm$^{-3}$.

\subsection{Lorentz factor evolution}\label{sec:lor}  

\begin{figure}[ht!]
\begin{center} 
\includegraphics*[width=0.45\textwidth]{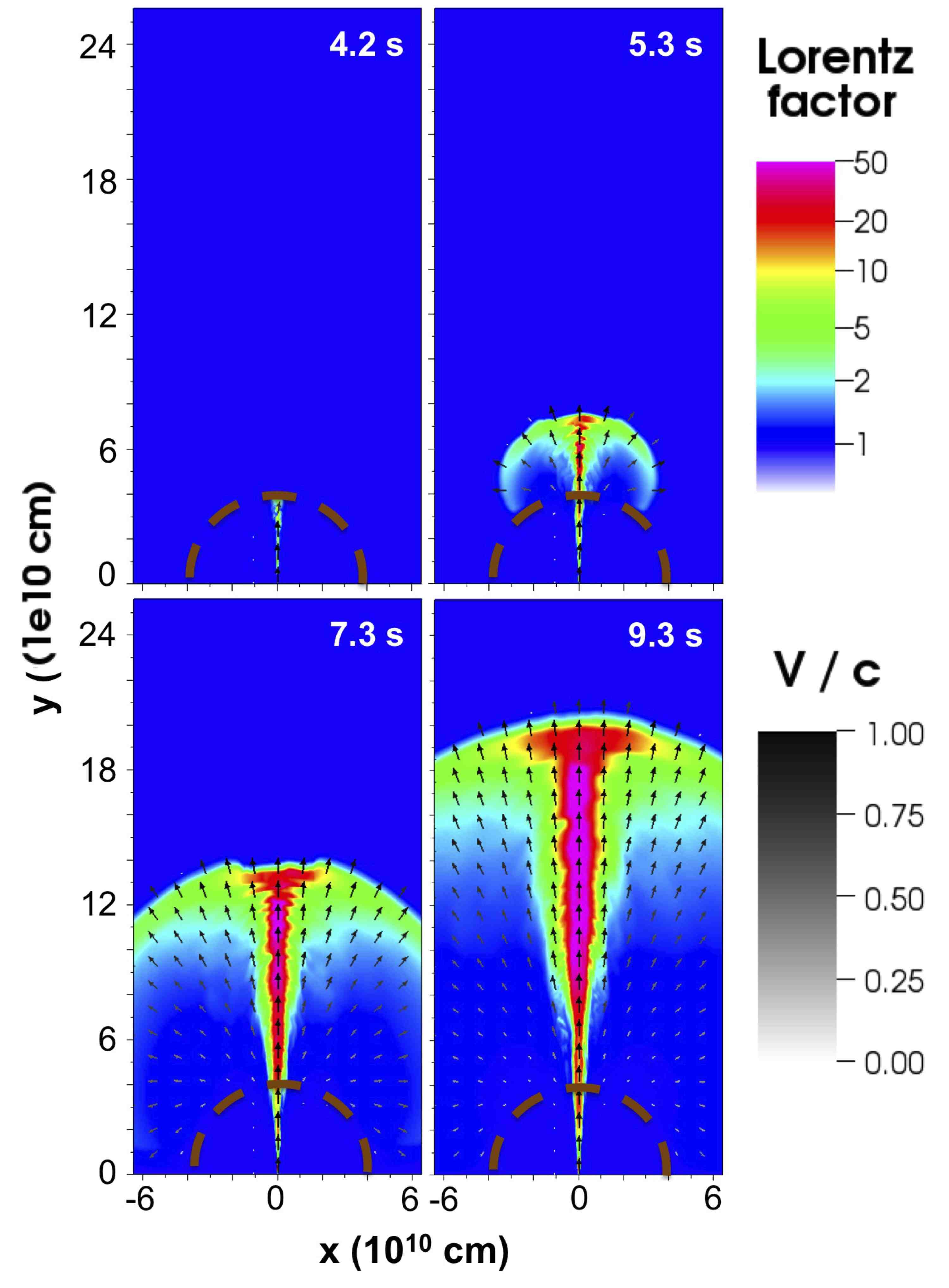}
\caption{Lorentz factor stratification maps and velocity field for
  different timeframes (4.2~s; 5.3~s; 7.3~s; 9.3~s) for model 3D
  LR. The brown dashed line indicates the stellar surface. A movie of this figure is available in http://www4.ncsu.edu/$\sim$dlopez/Simulations$\underline{\hspace{0.2cm}}$(published).html}
    \label{fig5}
\end{center} 
\end{figure}

\begin{figure}[ht!]
\begin{center} 
\includegraphics*[width=0.45\textwidth]{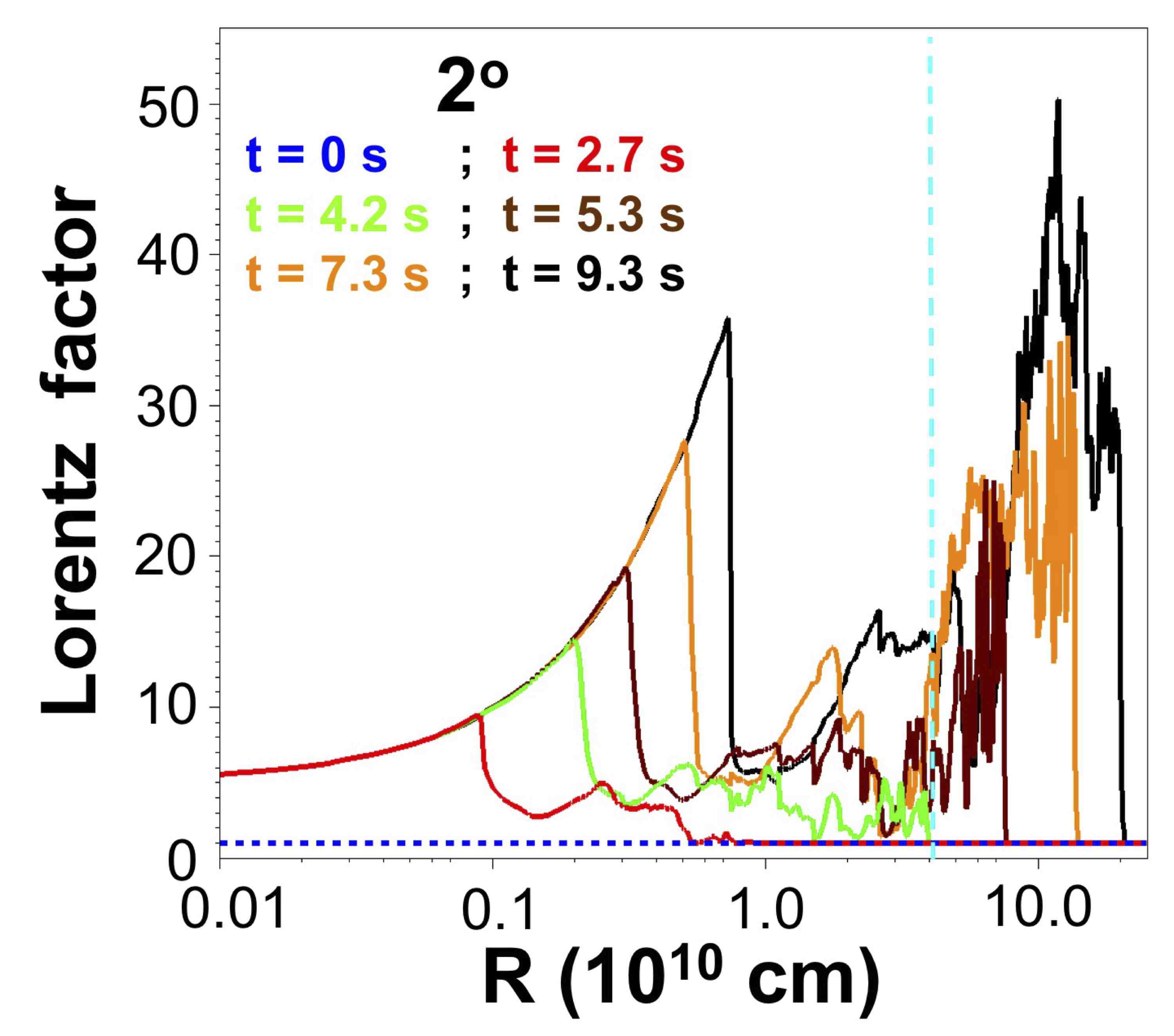}
\caption{Same as Figure~\ref{fig4} but for the Lorentz factor. A movie of this figure is available in http://www4.ncsu.edu/$\sim$dlopez/Simulations$\underline{\hspace{0.2cm}}$(published).html}
    \label{fig6}
\end{center} 
\end{figure}

In Figure~\ref{fig5} and Figure~\ref{fig6} we show the temporal
evolution for the Lorentz factor (with the velocity field also
present); and the radial Lorentz factor profile along the 2$^{o}$
radial path for model 3D LR. Before the breakout time only a
relativistic jet (with $\Gamma \sim$10) is present (see panel with
t=4.2~s from Figure~\ref{fig5}). In Figure~\ref{fig6} we see that the
SJ material for t$<$t$_{\rm{{bo}}}$ (blue, red and green lines)
reaches values close to $\Gamma=$15; and the UJ Lorentz factor remains
practically the same as the initial Lorentz factor
($\Gamma_{\rm{{0}}}$=5).
This behavior is consistent with what has already been seen in
previous GRB jet numerical simulations where the initial Lorentz
factor, prior to t$_{\rm{ {bo}}}$, reaches values close to 10
\citep{zwh04,miz06}. Once the jet breaks out of the stellar surface,
the jet is accelerated. The high internal energy is able to accelerate
material with Lorentz factors values of order $\Gamma \sim$100 in some
zones.  If accelerated with no energy dissipation, the jet's maximum
Lorentz factor would be 400 ($\Gamma_{\infty}$=$\Gamma_{\rm{{0}}}
\eta_{\rm{{0}}}$) \citep{mor07}.  In the panel with t=5.3~s from
Figure~\ref{fig5} (brown line in Figure~\ref{fig6}) we show how the
jet's forward shock and the recently formed cocoon produce a
``mushroom-like'' high-$\Gamma$ structure.  At later times
(t$>$t$_{\rm{ {bo}}}$) (orange, cyan and black lines in
Figure~\ref{fig6}) the mushroom-like structure grows bigger and its
Lorentz factor $\Gamma$ increases significantly.

\begin{figure}[ht!]
\begin{center} 
\includegraphics*[width=0.5\textwidth]{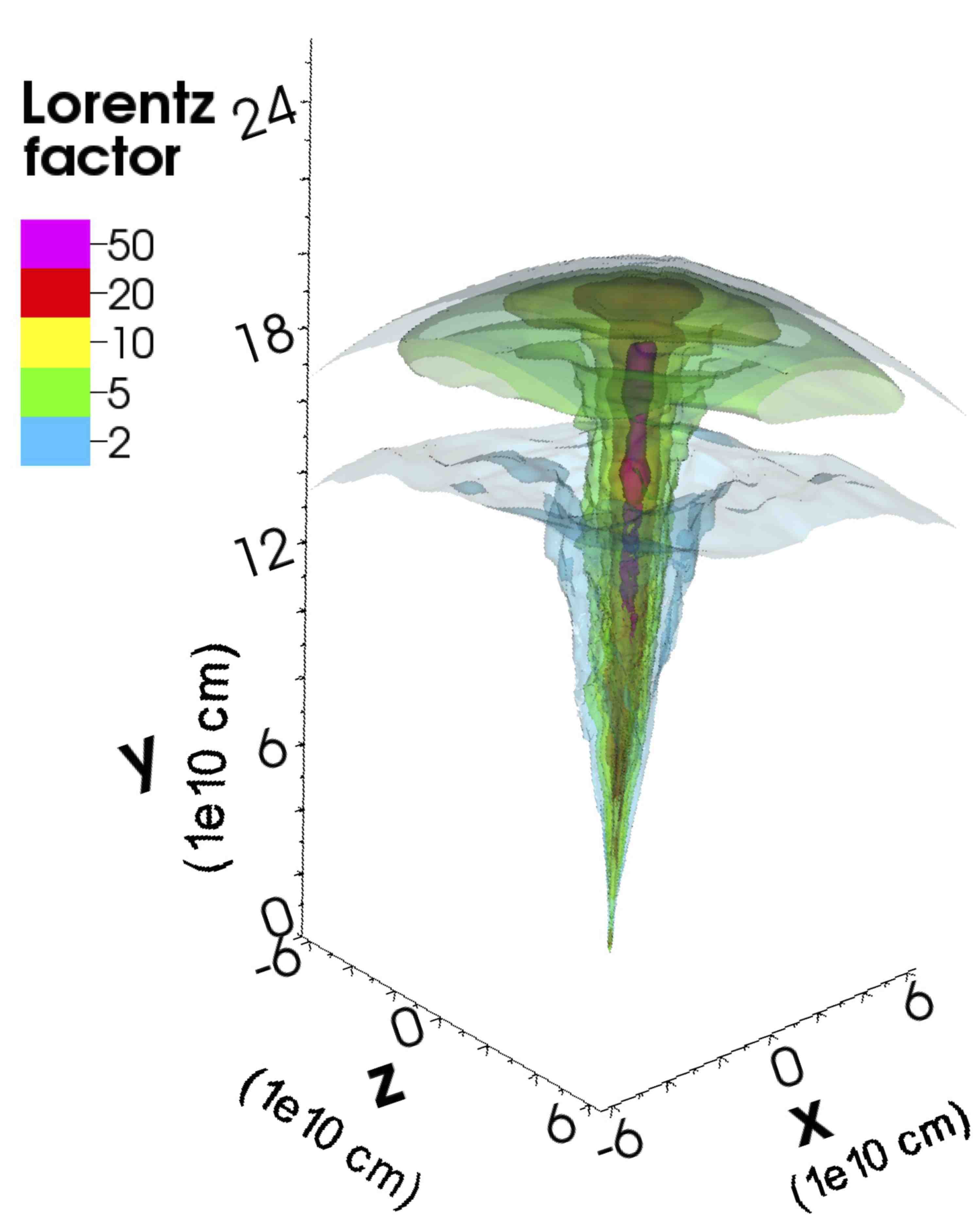}
\caption{Lorentz factor isocontour map at t=9.3~s for model 3D LR. The
  isocontour levels correspond to: 2, 5, 10, 20 and 50. A movie of this figure is available in http://www4.ncsu.edu/$\sim$dlopez/Simulations$\underline{\hspace{0.2cm}}$(published).html}
    \label{fig7}
\end{center} 
\end{figure}

To see the high Lorentz factor material in the jet, in Figure~\ref{fig7} we plot the $\Gamma$ isocontours for t=9.3~s. By this time the mushroom structure is evident and also certain regions
in the polar axis reach Lorentz factor values as high as $\Gamma=50$
(pink isocontours). Unfortunately our numerical domain does not permit
us to follow such high-$\Gamma$ regions and they escape the top
boundary after approximately 10 seconds. This is once more congruent
with the results from previous studies where after the t$_{\rm{
    {bo}}}$ the cocoon reaches values as high as $\Gamma \sim$15, and
the jet values of order $\Gamma \sim$100 \citep{miz06}. It must also
be noted that when the jet breaks out of the stellar surface, a
low-speed wind forms. This wind expands isotropically from the point
in the stellar surface where the jet drilled through, and moves at an
average speed vastly inferior (v$\le$0.01~c) to that of the jet.

\subsection{Resolution effects}\label{hrvslr}

\begin{figure}[ht!]
\begin{center} 
\includegraphics*[width=0.5\textwidth]{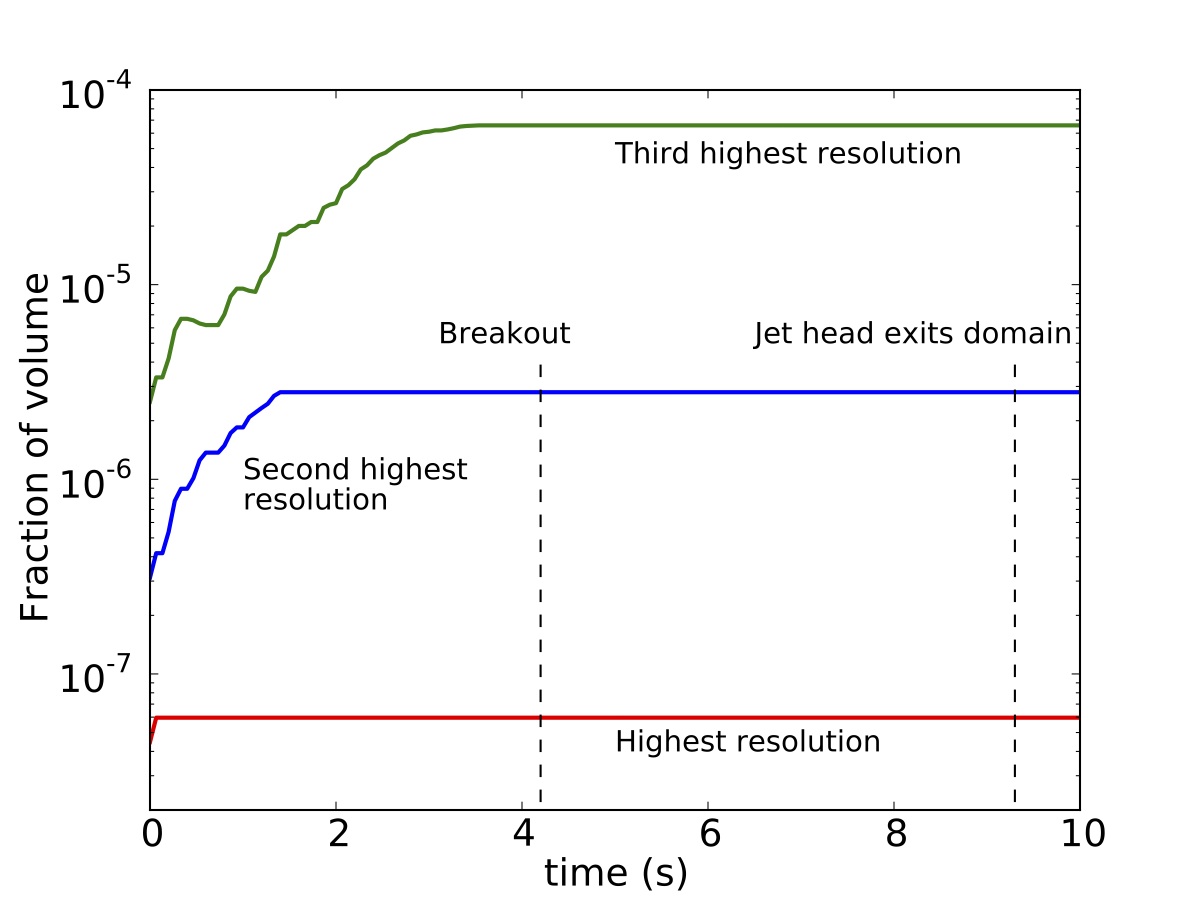}
\caption{Temporal evolution of the fraction of the volume that the two
  finest level occupied. The red line corresponds to the finest level
  ($\Delta \sim$10$^7$~cm), and the blue and green lines to the second
  and third finest ($\Delta \sim$10$^8$~cm).}
    \label{fig8}
\end{center} 
\end{figure}

In order to be able to evolve the initial setup up to integration
times of order $\sim$10~s and to resolve the jet-progenitor with a
suitably fine grid an AMR mesh was used. In Figure~\ref{fig8} we show
the fraction of the volume that the three finest resolution levels
occupied as a function of time. The finest grid level, the one with
which the base of the jet was resolved ($\Delta \sim$10$^7$~cm, red
line in Figure~\ref{fig8}) occupied less than 10$^{-7}$ of the entire
volume. Meanwhile, the two next finest levels, which followed the
propagation of the jet through the progenitor ($\Delta \sim$10$^8$~cm,
blue and green lines in Figure~\ref{fig8}), occupied less than
10$^{-6}$ and 10$^{-4}$ of the volume. Needless to say, if we had used
a fixed mesh with a comparable resolution to that with which the base
of the jet was resolved, it would have required $\sim$10$^6$ more
computational power (compared the computational power used in our
simulations), and thus the benefit from using an AMR scheme.

\begin{figure*}[ht!]
\begin{center} 
\includegraphics*[width=0.7\textwidth]{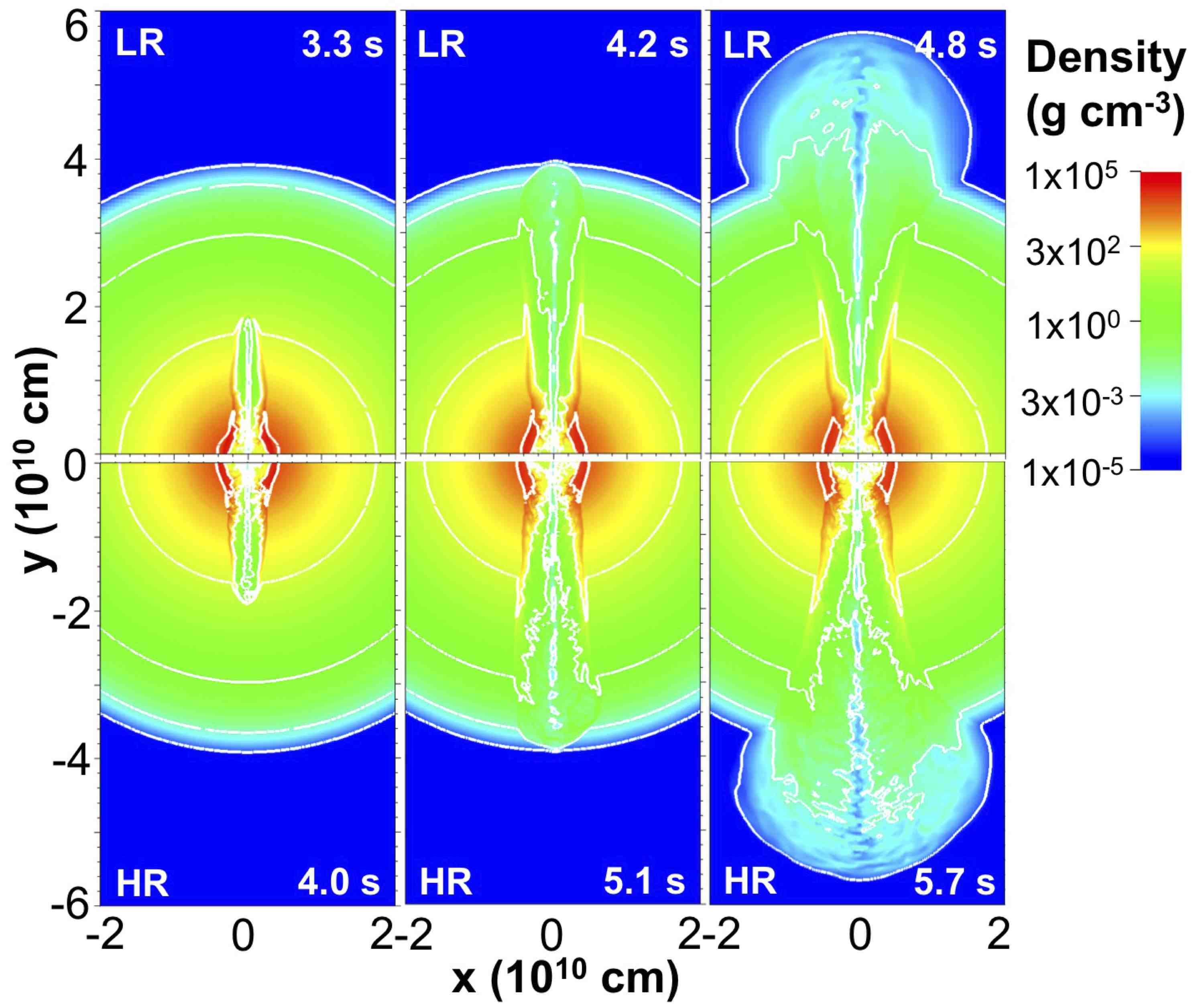}
\caption{Density stratification maps (g~cm$^{-3}$) for the 3D LR model
  (upper panels) and the 3D HR model (lower panels). For both models,
  we show representative timeframes from each of the two main phases
  (see text for discussion). The isocontour levels are the same as the
  ones indicated in Figure~\ref{fig1}.}
    \label{fig9}
\end{center} 
\end{figure*}

To verify that the evolution of the jet from our results is not
dependent on the numerical resolution, we ran a new model with the
same setup and physics but with a maximum resolution two times finer
than for the low resolution (see Section~\ref{sec:sims} for details).
In Figure~\ref{fig9} we show the density profiles for the 3D HR and 3D
LR case. In each case we show the two main phases already discussed in
Section \ref{sec:morph}: prior to the breakout phase; the breakout;
and the post breakout phase. We must note that the selected timeframes
for each of the phases were chosen arbitrarily so that the LR and HR
cases resemble each other, and hence their basic morphology
characteristics can be compared. The main result is that the basic
morphology characteristics from each phase (see discussion in
Sections~\ref{sec:morph}$-$\ref{sec:lor}), are well reproduced
independently of the numerical resolution.  Unlike the results from
the numerical 3D jet GRB study from \citet{wang08} where high
resolution models gave qualitatively different jet dynamics, we obtain
consistent jet behavior independently of the resolution.
 
Among the differences associated with the resolution, are a higher
level of turbulence and a slower advance of the jet head in the HR
model. The latter's jet moves $\sim$20\% slower than the LR case,
hence the breakout time for the HR case is t$_{\rm{ {bo}}}$=5.1~s. The
jets velocity resolution difference is due to the fact that the HR
case has a wider jet ($\sim$5\% wider than the LR case). Since we are
powering both jets equally, the narrow-LR jet will move faster.  The
turbulence resolution difference is due to the fact that the LR
simulation has higher diffusion, and thus suppresses the small scale
instabilities which are present in the HR model. The higher amount of
turbulence in the HR model also slows it down (compared to the LR
model), this due to the fact that a larger fraction of the energy is
converted into turbulence. Hence, reducing the HR jet's kinetic energy
and ram-pressure.  We must note that these two resolution effects are
consistent with what has already been seen in previous jet-collapsar
simulations, for example \citet{mor07}. Even though the latter study
is a two-dimensional one, it also presents more vortices in the HR
than in the LR case.

\begin{figure}[ht!]
\begin{center} 
\includegraphics*[width=0.45\textwidth]{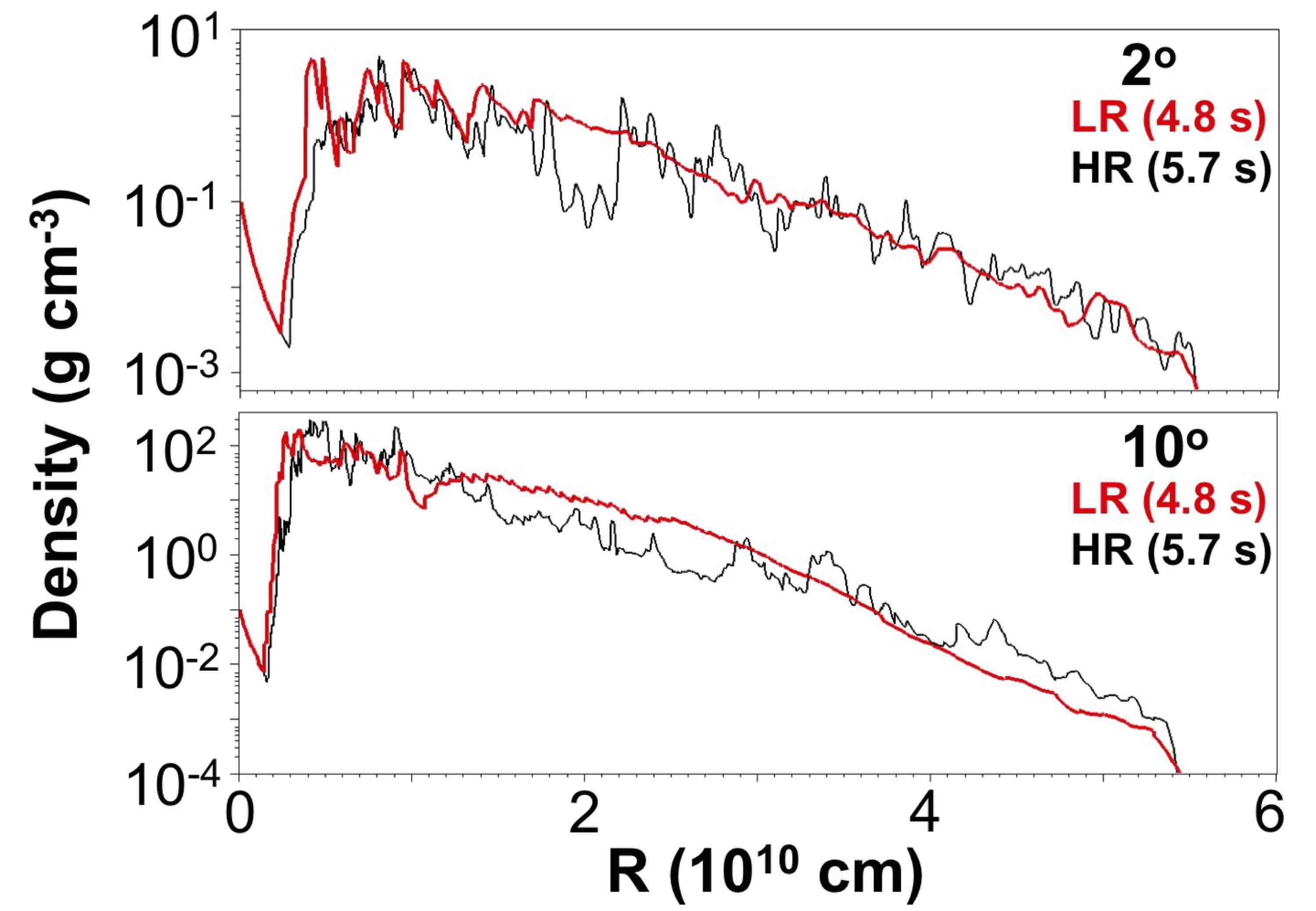}
\caption{Radial density profile (g~cm$^{-3}$) for the 3D LR model (red
  line) and the 3D HR model (black line). For both resolutions the
  2$^{ {o}}$ radial paths from the (+X,+Z) are shown in the upper
  panel, meanwhile the 10$^{ {o}}$ paths are shown in the lower
  panel.}
    \label{fig10}
\end{center} 
\end{figure}

To illustrate how in the HR model there is more variability than in
the LR one, in Figure~\ref{fig10} we plot the radial density profiles
for both resolution models. The upper panel of Figure~\ref{fig10} is
the radial density profile for a path within the jet (specifically a
2$^{ {o}}$ path), while the lower panel is the radial density profile
in the edge of the jet (10$^{ {o}}$ path). Inside the jet, there is no
major difference between the two resolution models. On the other hand,
at the edge of the jet the numerical resolution clearly affects the
density profile. Here the HR presents numerous depressions in the
density radial profile while the LR case has a radial profile which
follows a smoother distribution with less depressions and variability.

\begin{figure}[ht!]
\begin{center} 
\includegraphics*[width=0.5\textwidth]{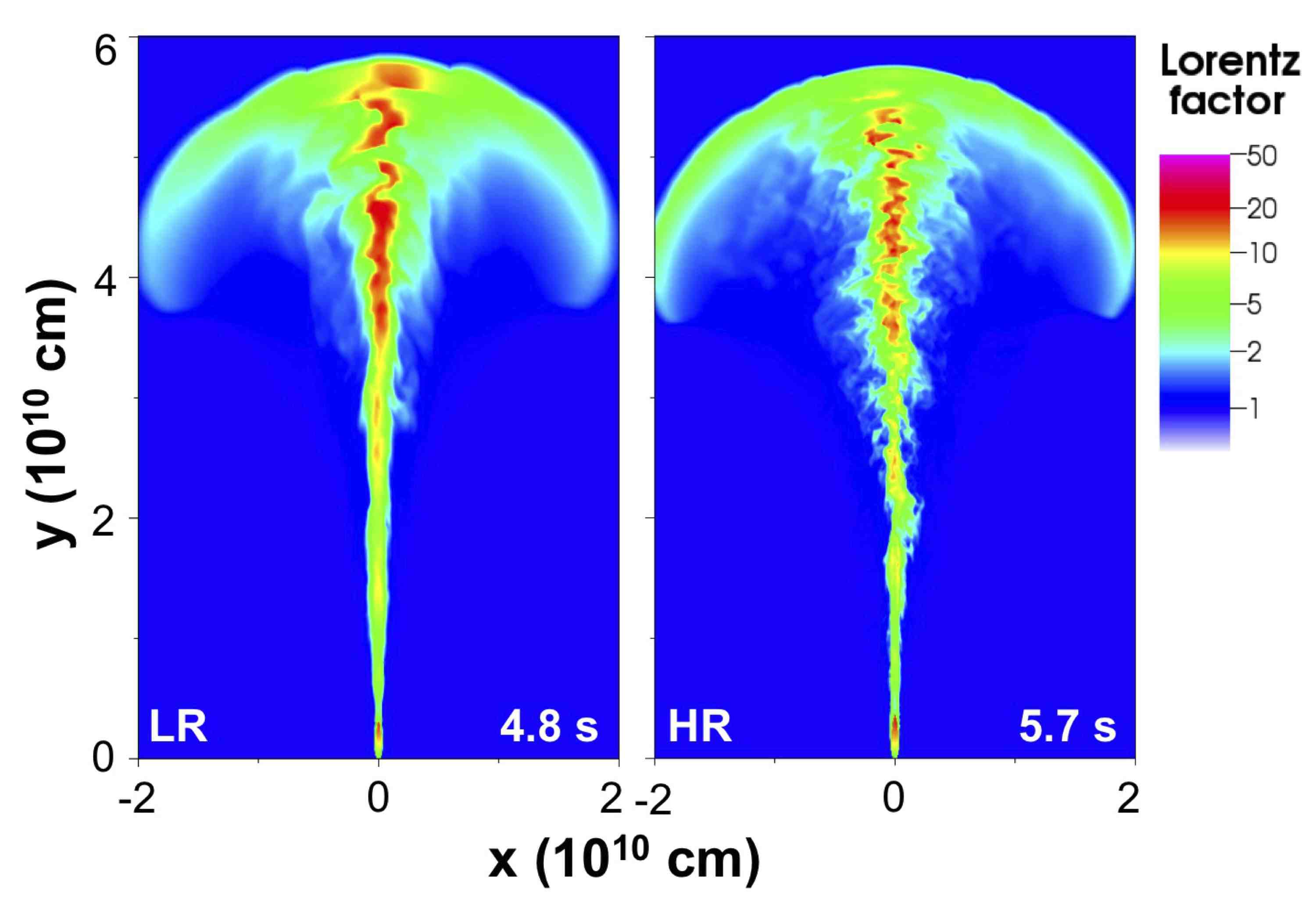}
\caption{Lorentz factor stratification maps for the 3D LR (left
  panel), and the 3D HR model (right panel). For both resolution
  models, we show a representative timeframe for when the jet has
  already broken out of the star and is moving across the ISM.}
    \label{fig11}
\end{center} 
\end{figure}

We also analyzed the effects of numerical resolution on the jet's
Lorentz factor distribution. Before the jet breaks out, apart from the
velocity with which the jet evolves, there is no clear difference
between the low and high resolution models. But at t$>$t$_{\rm{
    {bo}}}$ there are morphological changes due to the
resolution. Figure~\ref{fig11} shows the Lorentz factor map for the 3D
LR and the 3D HR models just after breaking out of the stellar
surface. The Lorentz factors are comparable but, as expected, the HR
case has more turbulent-like structures. For both resolutions the
low-density material (situated along the polar axis) has higher
Lorentz factors compared to the material in the edge of the jet. The
main difference, in qualitative terms, is that the high-$\Gamma$
regions from the LR model are split into smaller regions with lower
$\Gamma$ values in the HR model.

Another interesting aspect of our results is that the jet Lorentz
factor morphology resembles the precessing jet case from
\citet{zwh04}. In our simulations though, the zig-zagging pattern in
the Lorentz factor seems to be due to the ability of the high-$\Gamma$
material to wobble around the star's rotation axis and propagate
through paths of least resistance (see below for more details). The
observation of this effect is likely facilitated by the fact that we
have a larger dynamic range inside the star ($\delta_R = R_0 / R_{\rm{
    {min}}} = 41$) compared to the one from \citet{zwh04} ($\delta_R
=8.8$). Thus, there is enough range inside the star for 3D
instabilities to develop. In addition, the individual grid pixels in
the Zhang et al. (2004) simulations was not square. Rectangular pixels
generate more diffusion in the longest of the pixels direction, and
results are therefore not as robust as those from square pixel grid
simulations (where the diffusion is the same for all three
directions).

Finally, we must remark that we do not claim to reach convergency. If
we take the number of grid cells across the jet diameter as an
estimate of the Reynolds number (Re) of the simulation, we see that at
the present time we can only reach Re$\sim$200. Such Reynolds number
is approximately two orders of magnitude below the required Reynolds
number in which the jet behavior is independent of the resolution
\citep{be72}. Unfortunately, numerical simulations as those presented
in this study, with resolutions of at least two orders of magnitude
finer are not feasible (due to technical difficulties) to date.

\subsection{2D vs 3D simulations}

Finally, we checked how the evolution of the jet through the stellar
envelope varies in two and three dimensional simulations. For this we
ran an extra 2D numerical model with the same resolution
($\Delta$=1.25$\times 10^{8}$~cm), and the same parameter values
(luminosity, R$_{\rm{{i}}}$, $\theta_0$, $\Gamma_{\rm{{0}}}$, and
$\eta_0$, see section~\ref{sec:sims} for more details) as the 3D HR
case. Apart from the dimension difference from the 3D models, we
assumed polar axis symmetry in the 2D simulation; thus in reality we
only simulated half of the x-axis domain (i.e x$_{\rm{
    {min}}}$=0). 

\begin{figure*}[ht!]
\begin{center} 
\includegraphics*[width=0.9\textwidth]{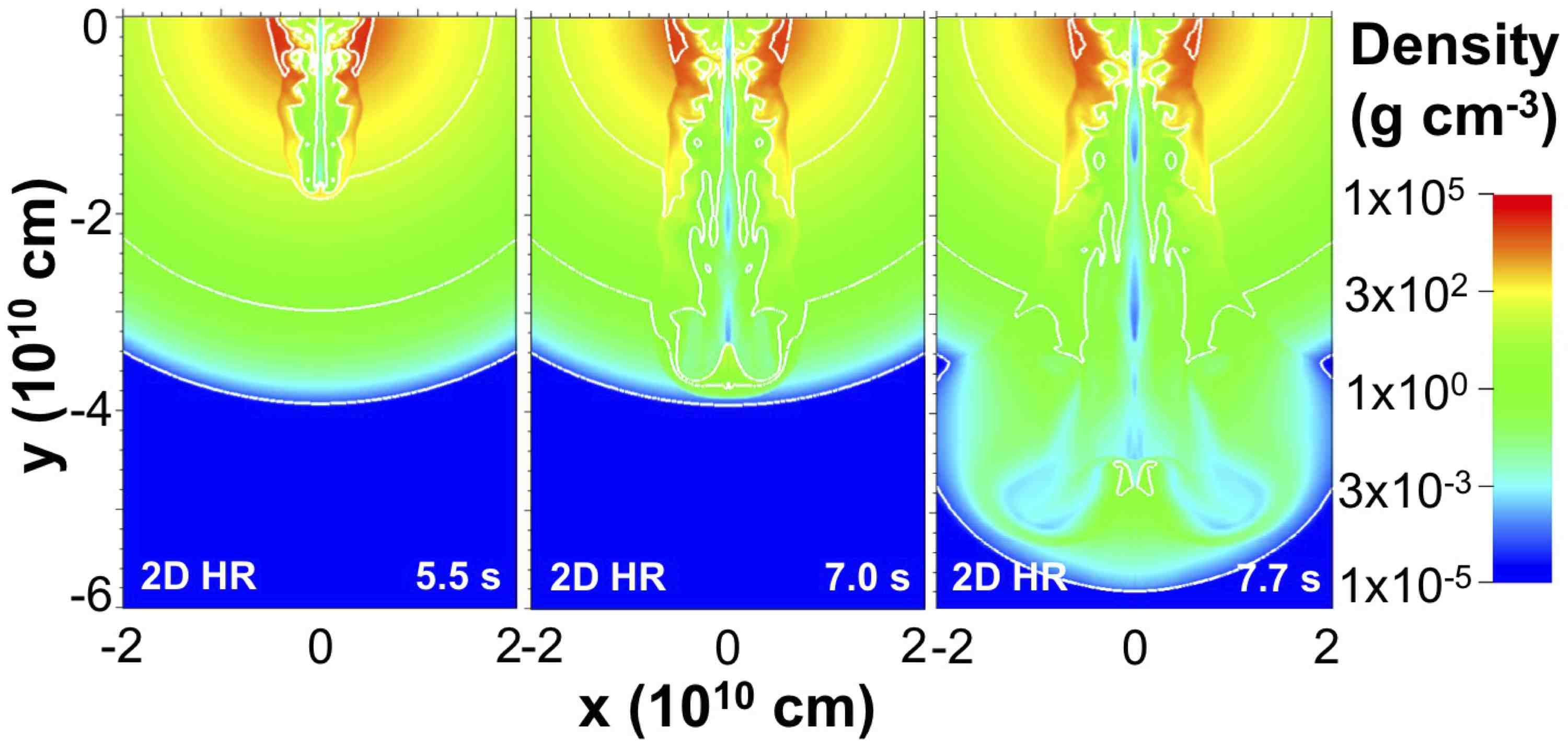}
\caption{Density stratification maps (g~cm$^{-3}$) for the 2D
  model. For this we show representative timeframes from each of the
  two main phases (see text for discussion). The isocontour levels are
  the same as the ones indicated in Figure~\ref{fig1}.}
    \label{fig12}
\end{center} 
\end{figure*}

The 2D simulation was carried out in cylindrical coordinates, with the polar axis coincident with the jet axis. In Figure~\ref{fig12} we show the density stratification maps for the 2D
model at the two phases (as well as for the breakout time):
a. t$<$t$_{\rm{ {bo}}}$; b. t$\approx$t$_{\rm{ {bo}}}$, and
c. t$>$t$_{\rm{ {bo}}}$). The timeframes for each of the 2D phases
shown in Figure~\ref{fig12} were chosen so that they resemble the
correspondent timeframes from the 3D HR case. The basic morphology in
the 2D case resembles that from the 3D model. In both cases we see a
collimated jet that manages to drill through the stellar
envelope. Apart from the polar axis symmetry imposed in the 2D model,
there are many subtle differences between the 2D and 3D results:

\noindent
1. The jet moves slower in the 2D model than in the 3D one (congruent
with \citet{zwh04}). Thus, the 2D breakout time is larger (t$^{
  {2D}}_{\rm{ {bo}}}$=7~s) than for its correspondent 3D HR model. The
reason for the slower jet's motion in the 2D model is the imposed
symmetry. Not only is the jet axis-symmetric, but also the stellar
material in front of the jet has to remain symmetric at all times. The
SJ material can only escape from the jets plug sideways, so a lot of energy
ends up going into accelerating this stellar material. In 3D models,
instead, the jet can deflect slightly and go around the plug (rather
than continuing to accelerate it, see below). Finally, the 2D jet
model has a SJ material mildly broader ($\approx$20\%) than its respective
from the 3D model.

\noindent
2. Even though the 2D model has the same resolution, the 2D jet
presents less turbulent-like morphology than what is present in the 3D
HR case. Also, the cocoon is less turbulent and broader in the 2D
scenario, as is the case in the numerical study of \citet{zwh04}.

\noindent
3. Two low-density plumes are present in the 2D simulations (see right
panel from Figure~\ref{fig12}). It must be noted that due to the
imposed axis-symmetry, the plumes actually correspond to a low-density
torus around the jet head (if it were a three dimensional domain and
not two dimensional). Such low-density torus is not present in any of
the 3D simulations. This is somewhat similar to the findings from
\citet{zwh04} where the head of the jet is noticeably different
depending on the simulations dimensionality (either 2D or 3D).

\noindent
4. As shown in the upper panel of Figure~\ref{fig14}, where we plot
the density profile along the polar axis for the 2D and 3D models, the
2D density radial profile is less turbulent and the depressions are
more profound than those in the 3D density radial profile.  Also, the
SJ material is mode dense by nearly two orders of magnitude in the 2D
scenario close to the jet head.

\begin{figure}[ht!]
\begin{center} 
\includegraphics*[width=0.45\textwidth]{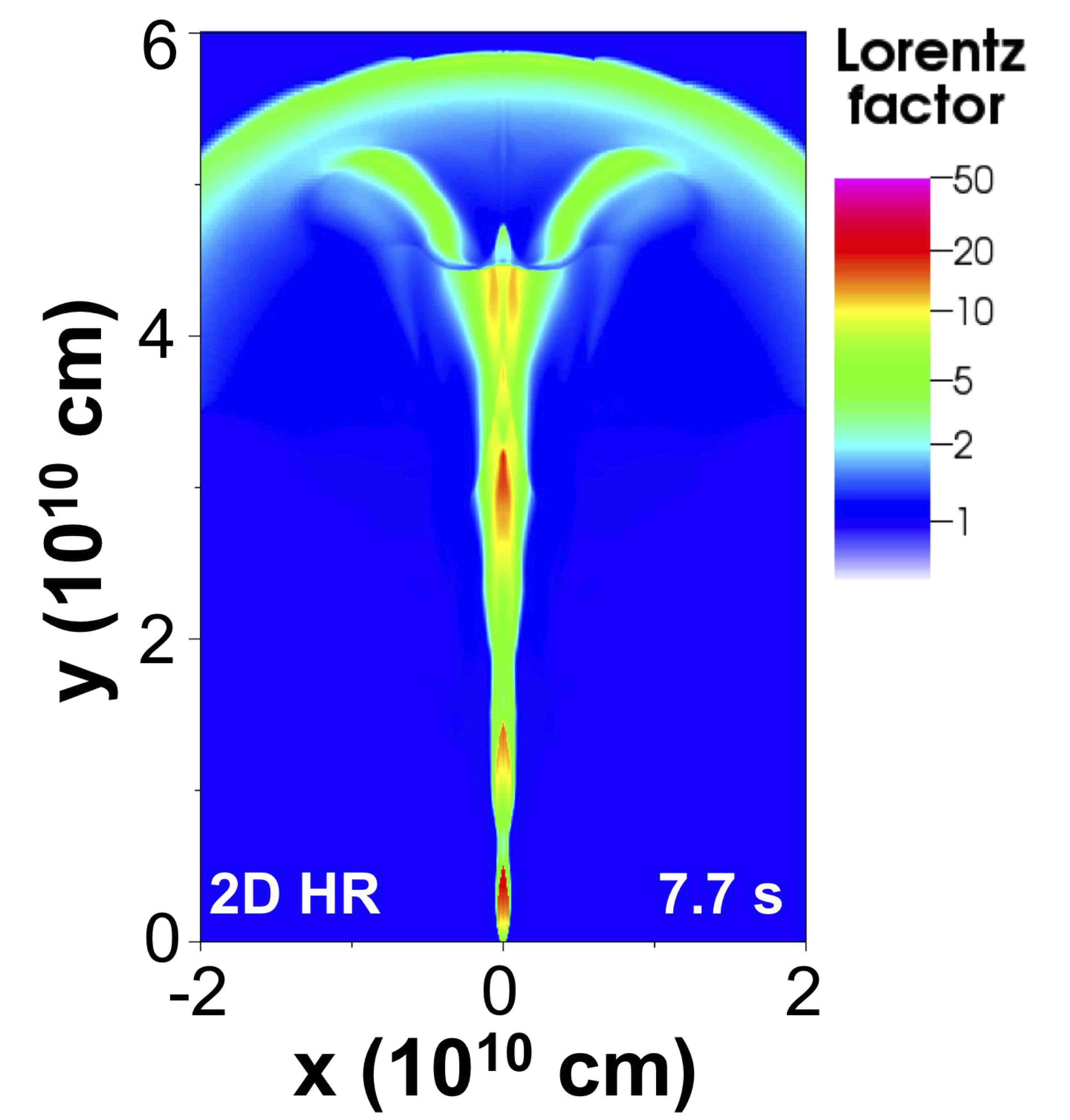}
    \caption{Lorentz factor stratification map for the 2D model at t=7.7~s.}
    \label{fig13}
\end{center}  
\end{figure}

\begin{figure}[ht!]
\begin{center} 
\includegraphics*[width=0.45\textwidth]{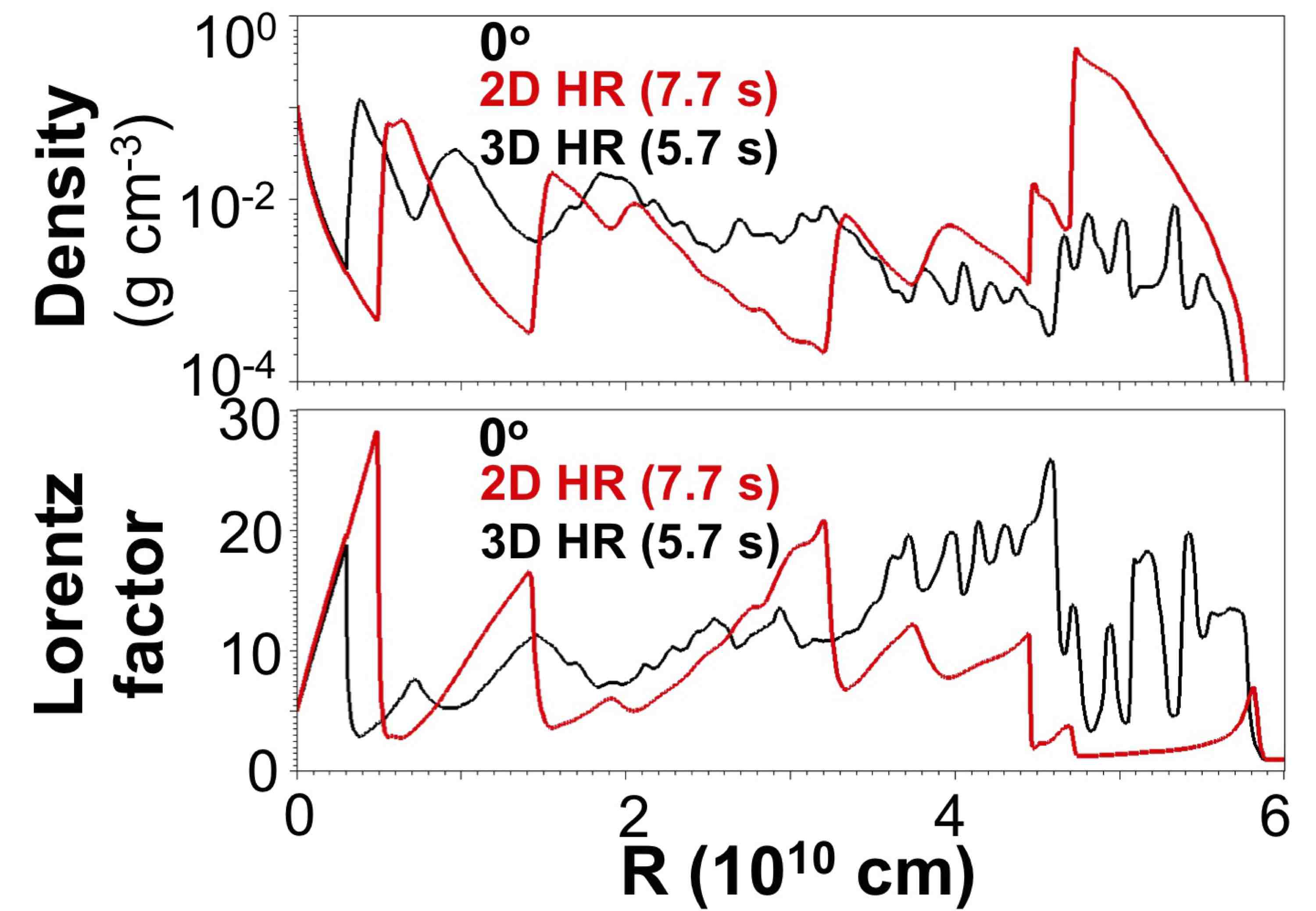}
    \caption{Radial density profile (g~cm$^{-3}$) (upper panel), and
      the radial Lorentz factor (bottom panel) for both the 2D model
      (red line) and the 3D model (black line) for the timeframe when
      the jet has broken out of the star. For both models the path is
      a polar axis (0$^{ {o}}$) radial paths from the (+X,+Z)
      quadrant.}
    \label{fig14}
\end{center} 
\end{figure}

In Figure~\ref{fig13} we show the Lorentz factor structure for the 2D
case (once the jet has just broken out of the stellar surface). Even
though the mushroom $\Gamma$ structure forms, the 2D Lorentz factor
morphology is noticeably different to that from the 3D HR case. The 2D
$\Gamma$ structure presents much less variability.

The 2D low-density regions have high $\Gamma$ values of up to 15-20,
values which are in agreement with those obtained by other 2D GRB jet
studies \citep{zwh04, nag11}. The 2D model also has a SJ which is
broader than that from the 3D case.  As was the case for the density
map in the 2D model, the head front of the cocoon has less
turbulent-like $\Gamma$ structures. In order to clarify this point, we
show the radial Lorentz factor profile (along the polar axis) in
Figure~\ref{fig14} (lower panel). Not only is a smoother radial
profile present in the 2D, but also the high-density low-$\Gamma$)
relationship is present. The SJ from the 2D case is approximately two
orders of magnitude more dense than from the 3D model, but has a
rather smaller $\Gamma$ value (of at most 5).

\begin{figure}[ht!]
\begin{center} 
\includegraphics*[width=0.45\textwidth]{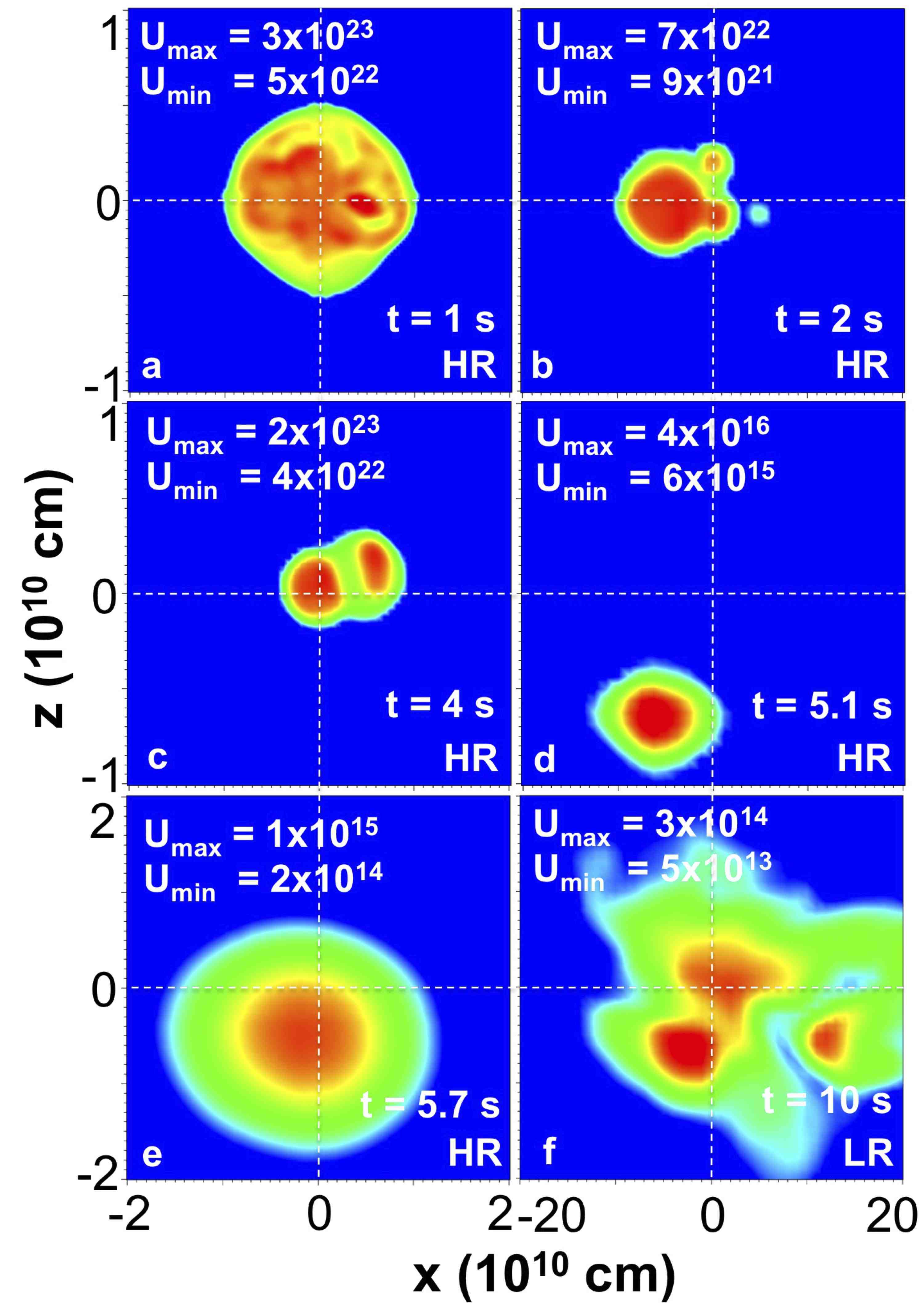}
\caption{Energy density (erg~cm$^{3}$) XZ stratification maps for the
  centroid of the head front of the jet-cocoon structure for different
  times. Panels a through e correspond to the 3D HR model, panel f
  correspond to the 3D LR model. Panel a. t=1s, b. t=2s, c. t=4s,
  d. t=5.1s, e. t=5.7s, f. t=10s; have the XY plane located at
  Y/(10$^{10}$~cm)=1.3, 3.3, 18.9, 39.9, 57.1 (respectively). In each
  panel the maximum and minimum values of the forward shock's energy
  density centroid (in erg~cm$^{3}$) is indicated. Notice how some
  panels have different scales.}
    \label{fig15}
\end{center} 
\end{figure}

In order to analyze how much the jet changes direction as it drills
through the progenitor star (and later through the ISM once the jet
has broken out of the star) in a three dimensional domain, we plot in
Figure~\ref{fig15} the energy density ($U$) map in the XZ plane. The XZ
planes shown for each timeframe correspond to the position where the $U$
centroid of the forward shock front was located at (see caption of
Figure~\ref{fig15} for more details). Panels a through d show the $U$
map for when the jet is drilling through the stellar progenitor. In
these it is noticeable how the centroid of the forward shock (CFS)
does not have a gaussian like profile (it may have turbulent like
behavior or even multiple spikes) and how the CFS wobbles around the
polar axis finding the spot of least resistance to proceed. For
example, notice how just before the jet breaks out of the star (panel
d), the CFS is located far from the polar axis
(x=-0.3$\times$10$^{10}$~cm , z=-0.6$\times$10$^{10}$~cm). Panel e
shows how once the jet has broken out of the star and the cocoon has
expanded thoroughly around the progenitor star, its correspondent CFS
also expands and also remains far from the polar axis.

\begin{figure}[ht!]
\begin{center} 
\includegraphics*[width=0.45\textwidth]{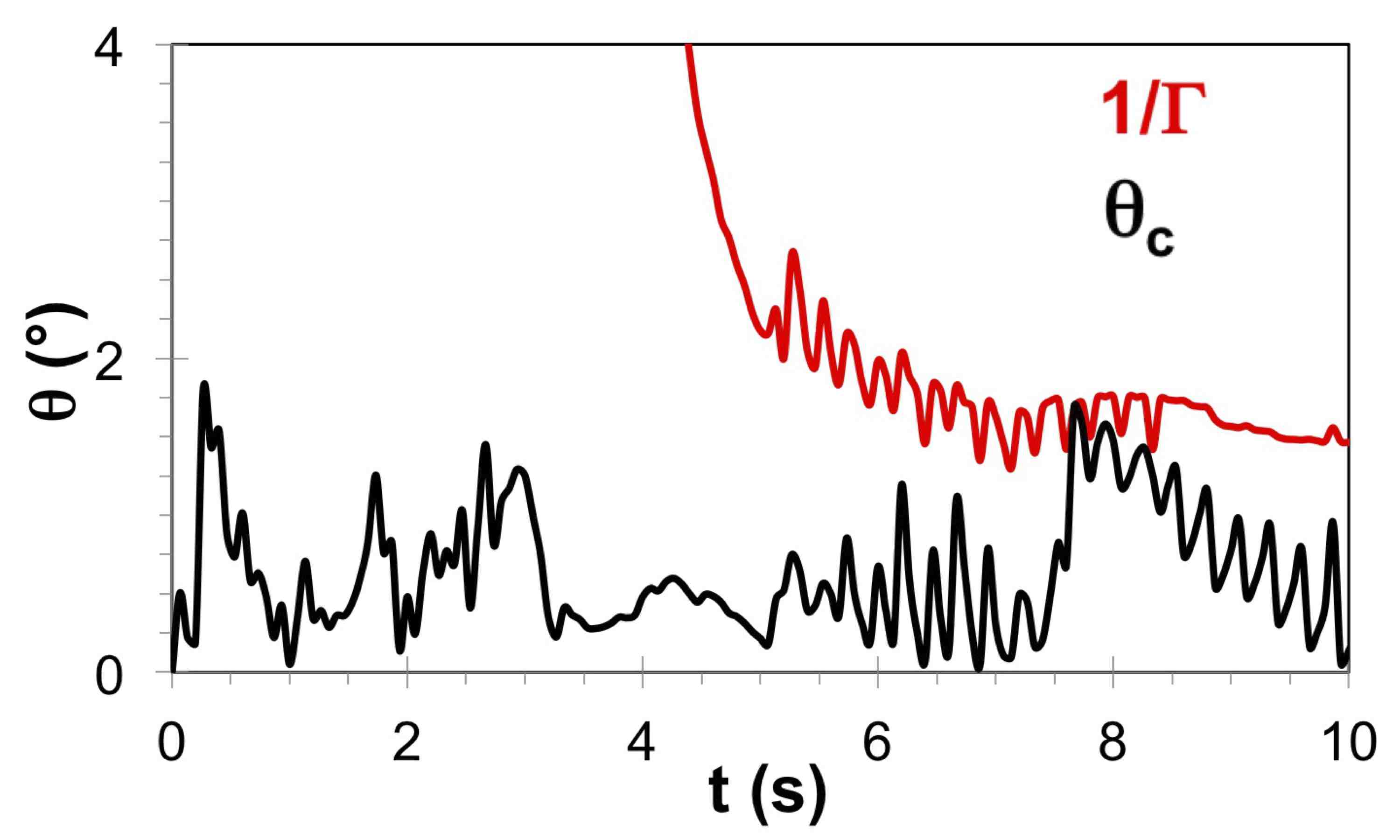}
\caption{Temporal evolution of the CFS (black line), and the
  relativistic collimation angle (red line).}
    \label{fig16}
\end{center} 
\end{figure}

To further understand the deflection of the jet inside the pre-SN
progenitor, we show the temporal evolution of the angle between the
CFS and the polar axis ($\theta$, black line in
Figure~\ref{fig16}). For a jet that is well aligned with the polar
axis then the CFS displacement angle would yield $\theta =0$, clearly
in Figure~\ref{fig16} this is not the case and the jet wobbles inside
the star (with $\theta$ oscillating between 0.1$^{o}$ and
2$^{o}$). Hence, the jet moves faster in 3D than in 2D because it is
able to wobble and move along the path with least resistance (apart
from having a narrower jet-cocoon). Note that $\theta$ is always
within the relativistic collimation angle ($1/\theta$, red line in
Figure~\ref{fig16}), thus the relativistic jet is causally connected at all times. 

\subsection{Limitations and comparison to other work}

As with all numerical work, the choices made in carrying out the
simulations reflect intentions and biases, and the current
investigation lacks in several aspects. For example, similar to
\citet{zwh04}, \citet{mor07, mor10}, and \citet{laz09} we assumed that
the star was static at all times which is clearly not the real case as
the pre-SN for long GRBs have very high angular momentum values
(J$>$10$^{15}$~cm$^2$~s) \citep{wh06}. We justify this by pointing out
that the dynamical timescale of the pre-SN is of order close to hours.
Then, since the integration time in our numerical simulations was of
order 10~s, we were safe to assume that the pre-SN progenitor remained
practically static at all times. In a previous study with a similar
setup \citep{laz11b} found that after $10^{2}$~s the pre-SN stellar
envelope had only expanded 2\% of its original size.

Another issue which can be improved is the ISM distribution. The
pre-SN progenitor that we use as the initial setup has no hydrogen
shell since during its stellar evolution it was lost by the presence
of a stellar wind (which will also affect the ISM surrounding the
pre-SN star).  So, in order to have full consistency the ISM should
have a density profile which was affected by the pre-SN wind, i.e. a
profile that follows a $\propto$R$^{-2}$ distribution \citep{zwm03,
  zwh04,can04,nag11}. But since the jet-cocoon system is an
ultra-relativistic flow, the density profile of the ISM will barely
affect the jet once this has just broken out of the stellar surface
\citep{mor07}. In fact, the GRB-jet needs to reach $\sim$10$^{14}$cm
for the ISM's profile to play a key role in the flow \citep{bm76,
  dc12}. Thus, we were secure to assume that the ISM density was
constant.

We use an adiabatic ($\gamma$=4/3) as our equation EOS \citep{zwm03,
  zwh04, miz06, mor07, mor10, laz09, nag11}. We do not take into
account the neutrino pressure, nor do we take into account the
gravitational effects from the central compact object. Even though it
has been shown that close to the pre-SN's progenitor nucleus the
neutrinos play an important role \citep{lc09}, since the inner
boundary was set so far away, R$_{\rm{{i}}} \sim$10$^9$cm, equivalent
to approximately $10^{4}$ gravitational radii, from the region where
neutrinos dominate (and where the compacts object relativistic effects
must be taken into consideration), the neutrino and relativistic
effects were safely ignored.

Since the follow up of newly formed elements was not the aim of this
study and that the calculation of such new elements would have not
permitted us to study the flow both at an adequate resolutions and for
the long times desired, nuclear burning was not included. Also, even
though magnetic fields will affect the emissivity of the jet
\citep{miz06}, and could even give origin to variability in the light
curve \citep{bh98}, they were disregarded due to the technical
difficulties when following a magnetized relativistic flow with an
adaptive mesh code.

\section{Conclusions}\label{sec:conc}

We present, for the first time, 3D AMR simulations of GRB jets
expanding inside a realistic pre-SN progenitor and then flowing
through the interstellar medium. Our numerical simulations, confirm
that relativistic jets can propagate and break out of the progenitor
star while remaining relativistic.

The morphology is divided into two main phases:\\
1. Pre-t$_{bo}$. During this phase the jet head moves at mildly
relativistic velocities ($\sim$c/2) inside the
progenitor's stellar envelope. \\
2. Post-t$_{bo}$. Once the jet breaks out of the surface, it
accelerates and reaches Lorentz factors of order $\Gamma \sim$~50.

The initial progenitor density profile is reshaped by the forward and
reverse shocks. The material between the forward and reverse shocks
break the two-dimensional symmetry in the numerical simulations.

We obtain similar behavior independently of the numerical
resolution. The resolution does not affect in great detail the flow
and the morphology in each phase is well reproduced. Still, the amount
of turbulence and variability observed in the simulations is higher
for higher resolutions. Also, for finer numerical resolutions the jet
moves slower; and regions with high Lorentz factors break up into
smaller regions with lower $\Gamma$ values.

The propagation of the jet head inside the progenitor star is slightly
faster in 3D simulations compared to 2D ones at the same
resolution. This behavior is due to the fact that the jet in 3D
simulations is narrower and can wobble around the jet axis finding the
spot of least resistance to proceed. Most of the jet properties, such
as density, pressure, and Lorentz factor, are only marginally affected
by the dimensionality of the simulations and therefore results from 2D
simulations can be considered reliable. If, instead, more detailed
properties such as variability are to be investigated, simulations carried
out in the proper dimensionality (i.e. 3D) are required.

\acknowledgments 
We thank S.E. Woosley and A. Heger for making their pre-SN models
available, and the referee for comments, suggestions and constructive
criticism which helped improve the original version of the manuscript.
The software used in this work was in part developed by the
DOE-supported ASC/Alliance Center for Astrophysical Thermonuclear
Flashes at the University of Chicago.  This work was supported in part
by the Fermi GI program grants NNX10AP55G and NNX12AO74G (D.L. and
D.L.-C.). B.J.M. is supported by an NSF Astronomy and Astrophysics
Postdoctoral Fellowship under award AST-1102796.



\end{document}